\documentstyle[preprint,aps]{revtex}
\begin{document}
\tightenlines
\title{Electron rescattering and the fragmentation dynamics of molecules in strong optical fields}
\author{F A Rajgara, M Krishnamurthy, and D Mathur}
\address{Tata Institute of Fundamental Research, 1 Homi Bhabha 
Road, Mumbai 400 005, India.}
\date{\today}
\maketitle
\begin{abstract}
We have probed the fragmentation dynamics in a bent triatomic molecule, water, a 
non-planar molecule, methanol, and a planar ring-structured molecule, benzene, 
using 100 fs duration pulses of linearly and circularly polarized, infrared, intensity-selected laser light. 
At laser intensities larger than 10$^{15}$ W cm$^{-2}$, the yield of singly and multiply charged atomic fragments from these molecules is suppressed when the light is circularly polarized. At lower intensities, the fragment ion yield is not significantly polarization dependent. This hitherto-unobserved intensity dependent effect of the polarization state of light on the fragmentation dynamics is rationalized using a simple electron rescattering 
model. Circular polarization switches ``off" electron rescattering and leads to suppression of multiple ionization and molecular fragmentation. Moreover, the degree of suppression is dependent upon the amount of energy transfer from the optical field to the molecule: the larger the energy transfer that is required for a particular fragmentation channel, the more marked is its suppression when circular polarization is used. 
\end{abstract}

\section{Introduction}

Ready availability of intense, pulsed infrared radiation from ultrafast lasers 
has opened new vistas for probing the nonlinear dynamics of atomic and molecular 
interactions in strong optical fields. Field-induced ionization of atoms and molecules 
is a foregone conclusion in such interactions. A special feature of strong-field 
ionization dynamics is that ionized electrons continue to ``feel" the effect of the 
optical field. The wavepacket that describes the ejected 
electron initially moves away from the vicinity of the parent. In the case of optical 
fields that are linearly polarized, the electronic wavepacket is pulled back towards the 
parent half a cycle after it was initially formed. The probability of recollision between 
the electron and the parent depends on the laser phase, and also the initial velocity 
and initial position of the electronic wave packet. Such 
rescattering allows the nuclear wavepacket to be probed with time resolutions that are 
lower than the pulse duration afforded by the laser that is used. The correlation between 
the electronic and nuclear wavepackets that are 
created in the ionization event has, recently, been utilized to probe the motion of the 
vibrational wavepacket of D$_2^+$ over several 
femtoseconds with unprecedented temporal accuracy of 200 attoseconds and spatial 
accuracy of 0.05 {\AA} \cite{niikura}.  Rescattering also 
affords other tangible benefits in that, in the case of atomic ionization, it 
gives rise to high harmonic emission \cite{corkum,krause}, generation of energetic 
electrons \cite{paulus}, multiple ionization \cite{weber} and attosecond pulse 
generation \cite{hentschel,paul}.  The effect of rescattering on diatomics like H$_2$ 
and D$_2$ has been probed \cite{niikura,alnaser} but its effect on the ionization dynamics is more difficult to 
discern because double ionization of such molecules occurs more readily through 
another strong-field process, enhanced ionization \cite{bandrauk,codling}. Recently, 
an intense-field, many-body S-matrix theory has been developed \cite{muth} that 
explicitly takes cognizance of electron wavepacket dynamics in determining 
ionization yields in polyatomic molecules. However, to the best of our knowledge, the 
effect of rescattering on ionization and fragmentation dynamics in molecules other than H$_2$ 
and D$_2$ has, hitherto, not been experimentally probed in systematic fashion. 
We report here results of experiments on the fragmentation dynamics of some polyatomic molecules using linearly and circularly polarized, femtosecond-duration, infrared, intensity-selected laser pulses of  
intensities that are large enough to generate electric fields 
of magnitudes that are comparable with the interatomic Coulombic fields. In the present study, 
we specifically explore the fragmentation dynamics in three types of molecules by 
utilizing the polarization properties of intense laser light over a range of peak 
intensities from 8$\times$10$^{14}$ to 10$^{16}$ W cm$^{-2}$. We have selected 
the molecular targets to typify a bent equilibrium geometry (water), a non-planar 
polyatomic (methanol) and a planar ring-structured polyatomic (benzene).

In going from linear to elliptically polarized light, it might be expected that the 
dynamics of the field-molecule interaction are influenced by one or more of the 
following factors: a) The trajectory of the electron (or electrons) ejected upon field-induced ionization (or multiple ionization); b) At the same laser intensity, the electric 
field amplitude is different for circular and linear polarization; and c) Angular 
momentum selection rules depend upon the polarization state of light. These factors 
manifest themselves, in the case of atoms, in changes in the ionization rate, changes in 
the energies of the ejected electrons, and on their angular distributions. In case of 
molecules, however, additional facets of the field-molecule interaction need to be 
considered, such as: (i) The polarization tensors in the molecule that might lead to 
alignment, specifically in the case of linear polarization; (ii) The dependence of the 
ionization rate on the angle between the induced dipole in the molecule and the 
electric field of the incident light; (iii) The rovibrational couplings in the electronic 
states that influence interatomic distances; (iv) The effect of enhanced ionization; and 
(v) Differences in the quantal descriptions of the electronic states that are excited, 
owing to the different angular momentum selection rules. All these parameters make 
for the difficulty of, and interest in the problem of understanding polarization-dependent 
molecular dynamics in intense light fields.

While controversies persist in theoretical formulations as to whether, and to what extent, the 
polarization state of the incident laser radiation might influence atomic ionization 
\cite{perelomov,klarsfeld,protopapas,patel}, experimental data with 
picosecond pulses appear to indicate that in both the intense (10$^{13}$ W cm$^{-
2}$) and super-intense ($>$10$^{15}$ W cm$^{-2}$) regimes, atomic ionization 
rates decrease with the ellipticity of the incident light. For instance, experiments on 
above-threshold-ionization have clearly shown that the ionization rate decreases with 
increase in ellipticity, and this has been rationalized by simple semi-classical 
formulations \cite{corkum2,dietrich,liang}. 
In the tunnel ionization regime, multielectron dissociative ionization of N$_2$ 
has been studied by 100-fs-long laser pulses of intensity in the 10$^{15}$ W cm$^{-2}$ range, 
using linearly and circularly polarized infra-red light \cite{banerjee,hering}. 
Substantial suppression of ionization channels has been observed in the case of 
circularly polarized light, even when laser intensities were appropriately adjusted to 
ensure that the laser field experienced by N$_2$ was identical in the two cases.  Interestingly, the enhanced ionization mechanism was shown to be valid for multiple ionization of N$_2$ with circularly polarized light \cite{banerjee}. Circularly 
polarized laser light, of intensity in the range 10$^{13}$-10$^{15}$ W cm$^{-2}$, 
has also been recently shown to lead to a reduced propensity for ionization of a chiral 
molecule \cite{chiral} in the picosecond regime. On the other hand, it has also been 
reported that the fragmentation pattern of molecules is not largely influenced by the 
laser polarization: Talebpour {\it et al.} \cite{talebpour} have recently shown that for 
intensities up to 10$^{15}$ W cm$^{-2}$ using femtosecond duration pulses, the 
fragmentation pattern in benzene, and relative ratios of fragment ion yields, are essentially similar for 
linear and circular polarization. 

The dynamics of fragmentation of a molecule in intense fields can be 
perceived to occur in two steps.  Firstly, the intense laser field ionizes the molecule. 
The ionization mechanism could be multiphoton ionization, tunnel ionization or 
over-the-barrier ionization, depending on the intensity of the interacting laser field. 
Secondly, the molecular ion, either singly- or multiply-charged, dissociates on the 
repulsive molecular ion potential energy surface, giving rise to energetic fragment ions. Before the 
molecular ion rolls down the excited potential surface, rescattering of electrons that are ionized 
but undergo oscillation under the influence of the intense laser field significantly 
affects the fragmentation of the molecules. Is this rescattering process 
more significant for larger polyatomics, like benzene? 

Quantitative, theoretical analysis of each of these steps, and prediction of their relative importance with  
change of laser intensity and molecular properties such as size, is not possible. 
In our study of fragment ion yields obtained upon irradiation of water, methanol, and benzene by intensity-selected laser fields, we find that at intensities larger than 1$\times$10$^{15}$ W cm$^{-2}$, the ionization yields of all the atomic fragment ions, and of the single and multiply charged molecular ions, are smaller with circularly polarized light than with linearly polarized light. We have chosen to concentrate only on atomic fragment ions that are produced 
in the interaction for the following reason. In the intensity regime that we probe, reliable comparison of 
molecular ion yields is very difficult as ionization occurs well in the saturation 
regime. Atomic fragment ions are more convenient probes in this respect as, for example, 
C$^{3+}$ from benzene is observed only for intensities greater that 
1$\times$10$^{15}$ W cm$^{-2}$, and saturation occurs at intensities that are much higher. In comparison, molecular analogs like C$_6$H$_6^+$ and C$_3$H$_3^+$ would be well into the saturation regime even at 10$^{14}$ W cm$^{-2}$. 
The results of our measurements also show 
that while fragment ion yields are more or less independent of the polarization state of the laser at lower 
intensities, the situation alters for intensities in excess of 10$^{15}$ W cm$^{-2}$. Here, the fragment 
ion yields are significantly lower for circularly polarized light. We rationalize these observations in 
terms of a simple electron rescattering model and suggest a propensity rule that gives some insight into the importance of electron rescattering in the fragmentation dynamics of molecular systems. 

\section{Experimental method}

Our experimental apparatus and methodology have been described recently 
\cite{banerjee2} and only those features that are most pertinent to the present study 
are mentioned in the following. Light pulses (of wavelength 806 nm) were obtained 
from a high-intensity, chirped pulse amplification, titanium-sapphire laser system 
operating at 10 Hz repetition rate. The laser light was focused using a biconvex lens, 
of 15 cm focal length, in an ultra-high vacuum chamber capable of being pumped 
down to a base pressure of 2$\times$10$^{-10}$ Torr. Our vacuum chamber was flooded with 
H$_2$O, CH$_3$OH, or C$_6$H$_6$ vapor (after degassing by means of several freeze-pump-thaw cycles in a clean, 
greaseless vacuum line) such that typical operating pressures were in the range of 6$\times$10$^{-8}$ Torr. Ions formed in the laser-molecule interaction were electrostatically extracted 
into a two-field, linear, time-of-flight (TOF) spectrometer. The polarization state of 
the light was varied by use of a half-wave (or quarter-wave) plate. The extent of 
elliptical polarization is defined by the ellipticity parameter, $\epsilon$=(Ex/Ey); in our 
experiments circular polarization implies an $\epsilon$-value of 0.9-1.0. 

Focal volume effects play a very important role in determining the ionization 
pattern observed using time-of-flight (TOF) spectrometers.  By using an aperture in 
the extracting plate of the TOF one can choose the extent of focal volume to be 
sampled, rather than sample the entire Rayleigh range.  For example, recently it has 
been shown with molecules like N$_2$ and CS$_2$ \cite{banerjee2} that intensity-selective 
and intensity-averaged TOF spectra differ from each other, since different intensity 
regions are ``seen" by the TOF spectrometer due to the spatial variation in intensity 
over the focal volume. Intensity-selected measurements are very important for the 
intensity regimes that we are probing so that the large ion counts from the low 
intensity region that could swamp the detector are avoided. We have conducted the 
present experiments in intensity-selective  mode by placing an aperture of 
5 mm in front of our TOF spectrometer. However to ensure that the collection 
efficiencies are not compromised, we applied very high extraction voltages such that  
the extraction fields were $\geq$250 V cm$^{-1}$; measurements of the fragment ion 
yield as a function of the extraction voltage were made to ensure unit collection 
efficiency, even for energetic atomic fragments like C$^{3+}$. 

\section{Results and discussion}

From an extensive set of mass spectrometric data on the interaction of intense light with water, methanol and benzene molecules, we present in the following that subset of data that pertains to the question: 
How does the polarization state of the intense laser radiation affect the 
fragmentation pattern when femtosecond pulses are used?  

Figs. 1 and 2 show fragment ion yields obtained upon irradiation of H$_2$O and CH$_3$OH at an intensity of 10$^{16}$ W cm$^{-2}$. We note that in both molecules, circular polarization results in a distinct suppression of fragment ion yields. Fig. 3 shows corresponding data for C$^{q+}$, $q$=1-3, fragment ions obtained from C$_6$H$_6$, and similar suppression with circular polarization is observed. Earlier studies on benzene, carried out using nanosecond and picosecond pulses, yielded overall fragmentation patterns that are similar to those observed in the present femtosecond measurements, although there are some differences in relative intensities (see \cite{bhardwaj,neusser}, and references therein).
The ``ladder switching" mechanism, together with its 
modifications \cite{neusser}, accounted for the fragmentation pattern in earlier long-pulse experiments. 
However, in the present experiments, since the laser pulses are of only 100 fs 
duration, it might be expected that the ladder switching mechanism is not likely to be 
applicable. Here, the fragmentation is likely to be induced by population of an 
electronic excited state of the molecular ion that possesses a repulsive potential 
energy surface, at least in the Franck-Condon region. The potential energy surface and 
its energy will, of course, be distorted, in some indeterminate fashion by the intense 
laser field. The fragment ions that are formed will depend on the nature of the field-
distorted state and on the minimum energy path in the multidimensional potential 
energy surface. 

Conventional ladder switching mechanisms demand a large increase of 
unimolecular dissociation rates with internal energy. Consequently, in the 
multiphoton ionization scenario, the precursor ion dissociates before there is time for 
additional photons to be absorbed. This is the rationale for the nonobservation of 
metastable multiply charged precursor molecular ions in long pulse experiments. The 
ion pairs observed by Bhardwaj {\it et al.} \cite{bhardwaj} in picosecond experiments invariably had atomic ions, C$^+$ or C$^{2+}$, as one of the constituents. In contrast, the present experiments yielded strong signals corresponding to long-lived molecular ions C$_6$H$_6^{2+}$ and C$_6$H$_6^{3+}$ in addition to atomic ions like C$^+$ and C$^{2+}$. 
We found that the propensity for producing multiply 
charged molecular ions was also distinctly lower with circularly polarized light as compared to that 
with linearly polarized light for intensities in excess of 10$^{15}$ W cm$^{-2}$. 

Our results on benzene apparently differ from those of 
Talebpour {\it et al.} \cite{talebpour}, who measured identical ion yields for both linear and 
circular polarization, also in the femtosecond regime. But, we note that the two sets of 
measurements were conducted at different laser intensities. Moreover, it is not clear 
whether Talebpour {\it et al.} employed an intensity-selective technique in their 
experiments. Absence of this would imply that their TOF spectrometer would access a 
wide intensity range covering 10$^{12}$-10$^{14}$ W cm$^{-2}$, with an enhanced propensity of ion 
collection from the lowest intensity regions within the focal volume. To probe the apparent differences further, we have made ion yield measurement for different laser intensities. Table 1 shows the relative ion yields for C$^{q+}$ fragments at different intensities: the yields are nearly same for intensities $<$10$^{15}$ W cm$^{-2}$, a value that corresponds to the maximum intensity used by Talebpour {\it et al.}. At this, and lower, intensities we find that there is virtually no polarization dependence in the fragmentation pattern, in accord with the findings of 
Talebpour {\it et al.}. However, at even a slightly enhanced intensity (such as a peak intensity value of 2$\times$10$^{15}$ W cm$^{-2}$), the fragmentation is found to become marginally lower in the case of circularly 
polarized light. The degree of suppression becomes much more pronounced as the laser intensity is increased. 

So, how does one account for the hitherto-unsuspected effect of laser intensity on the polarization dependence of molecular fragmentation. As noted earlier, for 
the very short pulses used in these experiments, ladder switching is not applicable as 
one can safely assume that the nuclear motion in all three molecules would be negligibly small over time periods of the order of 100 fs. We invoke electron rescattering in order to qualitatively explain the observed suppression in the 
fragment ion yield at higher laser intensities. As in the case of multiple ionization in atoms, we assume fragmentation of the molecular ion to be dominantly due to the rescattering of the ionized electrons in the 
presence of the laser field. We invoke the following chronology of events. Upon irradiation, the target molecule initially undergoes tunnel ionization when the field intensity is large enough. The ionized electron does not 
totally ``leave" the molecule, but interacts with it under the influence of both the 
Coulomb force and the laser field. At low values of laser field (corresponding to I = 10$^{14}$ W cm$^{-2}$) the Coulomb field has a large influence in determining the motion of the wavepacket that describes the ejected electron. On the other 
hand, at large fields (corresponding to I = 10$^{16}$ W cm$^{-2}$), the electric field of the interacting laser becomes 
comparable in magnitude to the Coulomb field and, therefore, exerts a much larger 
influence on the electron trajectories. To determine the influence of the interacting 
field on the motion of the ejected electron wavepacket, we made a model calculation for a hydrogen atom. We 
compute the electron trajectory by numerically solving the classical equation of 
motions. 

The equation of motion along the x-axis is
\begin{equation}
m {\partial^2 x \over \partial t^2} {\bf \vec{x}} = {e q \over r^2} {\bf \vec{x}} + e {\bf \vec{E}}  {\bf \vec{x}},
\end{equation}
where $e$, $m$ are the charge and mass of the electron, $q$ is the charge on the molecular ion, $r$ denotes the distance of the electron from the ion and $\vec{E}$ is the laser field. We numerically solve the differential equations of motion along all the ${\bf \vec{x}}$ ${\bf \vec{y}}$ ${\bf \vec{z}}$ directions iteratively, with a time grid of 0.01 a.u.. Fig. 4 depicts classical electron trajectories that we have computed for linear and circular polarization at two different laser intensities. 
At an intensity of 10$^{14}$ W cm$^{-2}$, the large Coulomb interaction ensures that electron trajectories for both  polarization states are very similar. This is depicted in the lower panel of Fig. 4. So, in the lower intensity regime, if the fragmentation is due to the dissociation of the molecular ion due to impact of rescattered electrons, the fragmentation yield would be expected to be more or less independent of the ellipticity of the laser field. At higher laser intensities, like 10$^{16}$ W cm$^{-2}$, the optical field becomes dominant, and the electron trajectories are very different for the two polarization states. While rescattering of the ejected electron is possible with linearly polarized light, it is absent in the case of circularly polarized light. So, one would expect the fragmentation channels that are due to rescattering to be switched off in the latter case. 

We note that at large intensity, the electron trajectories for linear polarized light depend on exactly when the electron wavepacket is created. If the initial position of the ejected electron lies on the $y$ = 0 line (when ${\bf \vec{E}}$ is parallel to the $x$-axis), then the electron would be expected to take part in rescattering. As the initial value of $y$ deviates from zero, the electron rescattering probability becomes small. 

So, for high laser intensities, the absence of rescattering in the case of circular polarization reduces the extent of molecular fragmentation. It appears reasonable to attribute the differences in fragmentation that are experimentally 
observed to be directly attributable to the change in electron rescattering probability. Our model calculations are simple but demonstrative. However, they pertain to an atomic target. This simplicity begs the question: does molecular structure play a role in determining the overall strong field fragmentation dynamics?

In order to probe this, and to lay the groundwork for proper theoretical treatment, we consider in Fig. 5 how the suppression of fragmentation depends the appearance energy of fragment ions from specific parent molecules. The ion appearance energy is a measure of the ionization energy of the given fragment, say C$^{2+}$, plus the bond dissociation energy. The latter accounts for molecular structure effects and, hence, results in different values of appearance energy for C$^{2+}$ from benzene and methanol precursors. The appearance energy is, therefore, a measure of the amount of energy transfer from the optical field to the molecule that is necessary in order to produce a given fragment ion. Data in Fig. 5 demonstrate that circular polarization (the switching ``off" of electron rescattering) results in distinctly more marked suppression of fragmentation channels that require the largest energy transfer. 

\section{Summary and concluding remarks}

We have conducted experiments on intense-field dissociative ionization of water,  
methanol, and benzene vapor with linearly and circularly polarized laser light. We observe a distinct 
lowering of the propensity to produce multiply charged fragment ions from all these molecules 
when circularly polarized light is used at laser intensities in excess of 10$^{15}$ W cm$^{-2}$. At 
peak laser intensities lower than this, light-induced fragmentation appears to be 
more or less independent of the polarization state of the incident intense light.  The lowering of 
multiply charged fragment ion yields with circularly polarized light is attributed to 
the lowered probability of the rescattered electrons inducing dissociative ionization.

Our data also indicates that molecular structure effects are important in determining the degree of suppression that can be achieved by changing the polarization state of the incident laser radiation from linear to circular. Those fragmentation channels that require the largest transfer of energy from the optical field to the molecule are suppressed most markedly by using circularly polarized light; the suppression is less marked for those channels that require smaller amounts of energy transfer. 

The present set of experiments have probed electron rescattering from molecules more complex than diatomic species and have revealed new facets of strong-field phenomena that have hitherto not been considered. Both the intensity dependence of the suppression that has been observed as well as dependence on energy transfer will have to be accounted for in development of theoretical insights into molecular fragmentation dynamics in strong optical fields. 

Within the context of atomic ionization, Lambropoulos \cite{Lambropoulos} pointed out, thirty years ago, that the effect of light polarization on the multiphoton ionization of atoms is related, in a general sense, to the effect of field correlations \cite{field} of multiphoton processes. Both effects arise from the fact that the vectors of the radiation field affect, in nonlinear fashion, the transition amplitudes for multiphoton processes. The nonlinearity in the amplitude of the radiation field leads to ionization rates that depend on the correlation functions of the field, and not just on the absolute value of the field amplitude. When the circular polarization vector $\epsilon_x \pm i\epsilon_y$ is inserted in the expression for the transition amplitude, cross products of matrix elements involving the orthogonal components $\epsilon_x$ and $\epsilon_y$ occur, and these lead to the dependence of the ionization rate on the polarization state of the incident light field.  However, the dependence of polarization effects upon the intensity of the applied light field that has been observed in our experiments on water, methanol, and benzene remains unexplained within the framework of the prevailing wisdom that has been articulated above in simple terms. 

Within the framework of tunnel ionization, the ADK formalism \cite{ADK} sheds some light on how atomic ionization rates depend on the polarization state of the incident light. The ADK theory predicts that the ratio of ionization rate for circular polarization ($w_{circ}$) to that for linear polarization ($w_{lin}$) is
\begin{equation}
w_{circ}/w_{lin} = (\pi q^3/En^{*3})^{1/2},
\end{equation}
where, as before, $q$ represents the ionic charge state, $E$ is the electric field amplitude, and $n^{*}$ is the effective principle quantum number that is stipulated in the ADK formalism. This expression predicts a suppression of ionization probabilities in the case when linearly polarized light is replaced by circularly polarized light of the same field strength. Moreover, such suppression is expected to have a $I^{-1/4}$ dependence on laser intensity. While our results do not replicate the exact functional dependence on $I$, the suppression that is observed by us is, at first sight, accounted for within the ADK picture. Where the ADK picture fails is in accounting for the apparent threshold of 10$^{15}$ W cm$^{-2}$ that we observe for such suppression.  The ADK picture cannot, of course, be expected to account for specifically {\em molecular} effects that we have discovered here, such as the dependence of suppression on fragment ion appearance energy. 

\section{Acknowledgment}
Some of the preliminary measurements on H$_2$O were made by Gautam V Soni, and we thank him for the diligence and care with which these initial experiments were conducted. Our T$^4$ (TIFR Table-top Terawatt) laser system was partially financed by the Department of Science and Technology for which we are also grateful. We benefitted from the stimulating discussions of the benzene results with Haruo Shiromaru. 

\begin{table}
\caption{ Relative ion yields of C$^{q+}$ with respect to C$^+$ in the fragmentation of benzene at different laser intensities.} 
\begin{tabular}{lcc}
%\hline

Intensity & ~C$^{2+}$ & ~C$^{3+}$ \\
  \hline\hline
$<$10$^{15}$Wcm$^{-2}$ & ~1.1 & ~0.8 \\
 2$\times$10$^{15}$Wcm$^{-2}$ & ~1.3 & ~1.2\\
 8$\times$10$^{15}$Wcm$^{-2}$ & ~1.6 & ~5.2 \\
  \hline
\end{tabular}
\end{table}

\begin{figure}
\caption{Polarization dependence in the relative fragment ion yields for H$_2$O at 10$^{16}$ W cm$^{-2}$. The H$_2$O$^+$ molecular ion yields obtained with both polarization states were normalized to the same value in order to determine the relative fragment ion yields.}
\end{figure}

\begin{figure}
\caption{Polarization dependence in the relative fragment ion yields for CH$_3$OH at 10$^{16}$ W cm$^{-2}$. The CH$_3$OH$^+$ molecular ion yields obtained with both polarization states were normalized to the same value in order to determine the relative fragment ion yields.}
\end{figure}

\begin{figure}
\caption{Polarization dependence in the relative fragment ion yields for C$_6$H$_6$ at 10$^{16}$ W cm$^{-2}$. The C$_6$H$_6^+$ molecular ion yields obtained with both polarization states were normalized to the same value in order to determine the relative fragment ion yields.}
\end{figure}

\begin{figure}
\caption{a) Electron trajectories for ionization of H by linearly (solid line) and circularly (dashed line) polarized light of intensity 1$\times$10$^{16}$ W cm$^{-2}$. The position of the H-atom at the origin is indicated by X. All distances are indicated in atomic units (a.u.). The vertical axis defines the x-direction (see text) while the horizontal axis is the y-direction. The arrows indicate the classical motion of the ejected electron. b) Electron trajectories for ionization of H by linearly (solid line) and circularly (dotted line) polarized light of intensity 1$\times$10$^{14}$ W cm$^{-2}$.}
\end{figure}

\begin{figure}
\caption{Ratio of ion yields obtained with linearly and circularly polarized light for different fragment ions as a function of fragment ion appearance energy. The laser intensity was 10$^{16}$ W cm$^{-2}$. The solid line is to guide the eye.}
\end{figure}


\begin{references}

\bibitem{niikura}H. Niikura, F. Legare, R. Hasbani, M. Yu. Ivanov, D. M. 
Villeneuve, and P. B. Corkum, Nature (London) {\bf 421}, 826 (2003).
\bibitem{corkum}P. B. Corkum, Phys. Rev. Lett. {\bf  71}, 1994 (1993).
\bibitem{krause}J. L. Krause, K. J. Schafer and K. C. Kulander, Phys. Rev. Lett. {\bf 
68}, 3535 (1992).
\bibitem{paulus}G. G. Paulus, W. Nicklich, H. Xu, P. Lambropoulos, and H. 
Walther, Phys. Rev. Lett. {\bf 72}, 2851 (1994).
\bibitem{weber}Th. Weber, H. Giessen, M. Weckenbrock, G. Urbasch, A. Staudte, 
L. Spielberger, O. Jagutzki, V. Mergel, M. Vollmer, and R. D\"orner, Nature 
(London) {\bf 405}, 658 (2000).
\bibitem{hentschel}M. Hentschel, R. Kienberger, Ch. Spielmann, G. A. Reider, 
N. Milosevic, T. Brabec, P. Corkum, U. Heinzmann, M. Drescher, and F. Krausz, 
Nature (London) {\bf 414}, 509 (2001).
\bibitem{paul}P. M. Paul, E. S. Toma, P. Breger, G. Mullot, F. Auge, Ph. Balcou, H. 
G. Muller, and P. Agostini, Science {\bf 292}, 1689 (2001).
\bibitem{alnaser}A. S. Alnaser, T. Osipov, E. P. Benis, A. Wech, C. L. Cocke, X. M. Tong, and C. D. Lin, private communication.
\bibitem{bandrauk}A. D. Bandrauk, Coments At. Molec. Phys. D {\bf 1}, 97 (1999).
\bibitem{codling}K. Codling and L. J. Frasinski, J. Phys. B {\bf 26},  783 (1993).
\bibitem{muth}J. Muth-B\"ohm, A. Becker, S. L. Chin, and F. H. M. Faisal, Chem. Phys. Lett. {\bf 337}, 313 (2001).
\bibitem{perelomov}A. M. Perelomov, V. S. Popov, and M. V. Terentev, Sov. Phys. 
JETP {\bf 23}, 924 (1966).
\bibitem{klarsfeld}S. Klarsfeld and A. Maquet, Phys. Rev. Lett. {\bf 29}, 79 (1972). 
\bibitem{protopapas}M. Protopapas, D. G. Lappas and P. L. Knight, Phys. Rev. Lett.  
{\bf 79}, 4550 (1997).
\bibitem{patel}A. Patel, M. Protopapas, D. G. Lappas and P. L. Knight, Phys. Rev.  A 
{\bf  58}, R2652 (1998).
\bibitem{corkum2}P. B. Corkum, N. H. Burnett and F. Brunel, Phys. Rev. Lett. {\bf 
62}, 1259  (1989).
\bibitem{dietrich}P. Dietrich, N. H. Burnett, M. Ivanov and P. B. Corkum, Phys. Rev. 
A {\bf 50}, R3585 (1994).
\bibitem{liang}Y. Liang, A. Augst, M. V. Ammosov, S. Lazarescu and S. L. Chin, J. 
Phys. B {\bf 28}, 2757 (1995).
\bibitem{banerjee}S. Banerjee, G. R. Kumar and D. Mathur, Phys. Rev.  A {\bf 60}, 
R25 (1999).
\bibitem{hering}Ph. Hering and C. Cornaggia, Phys. Rev.  A {\bf 59}, 2836 (1999). 
\bibitem{chiral}M. Krishnamurthy and D. Mathur, Phys. Rev.  A {\bf 61}, 63404 
(2000).
\bibitem{talebpour}A. Talebpour, A. D. Bandrauk, K. Vijayalakshmi and S. L. Chin, 
J. Phys. B {\bf 33}, 4615 (2000).
\bibitem{banerjee2}S. Banerjee, G. R. Kumar and D. Mathur, J. Phys. B {\bf 32}, L305 (1999); {\it ibid.} 4277.
\bibitem{bhardwaj}V. R. Bhardwaj, K. Vijayalakshmi and D. Mathur, Phys. Rev.  A {\bf 59}, 1392 (1999).
\bibitem{neusser}H. J. Neusser, U. Boesl, R. Weinkauf and E. W. Schlag, Int. J. Mass Spectrom. 
Ion Processes {\bf 60}, 147 (1984). 
\bibitem{Lambropoulos}P. Lambropoulos, Phys. Rev. Lett.  {\bf 28}, 585 (1972), {\it ibid.} {\bf 29}, (1972) 453.
\bibitem{field}P. Lambropoulos, Phys. Rev. {\bf 168}, 1418 (1968), B. R. Mollow, Phys. Rev. {\bf 175}, 1555 (1968), G. S. Agrawal, Phys. Rev. A {\bf 1}, 1445 (1970).
\bibitem{ADK}M. V. Ammosov, N. B. Delone, and V. P. Krainov, Sov. Phys. JETP {\bf 64}, 1191 (1987).

\end{references}
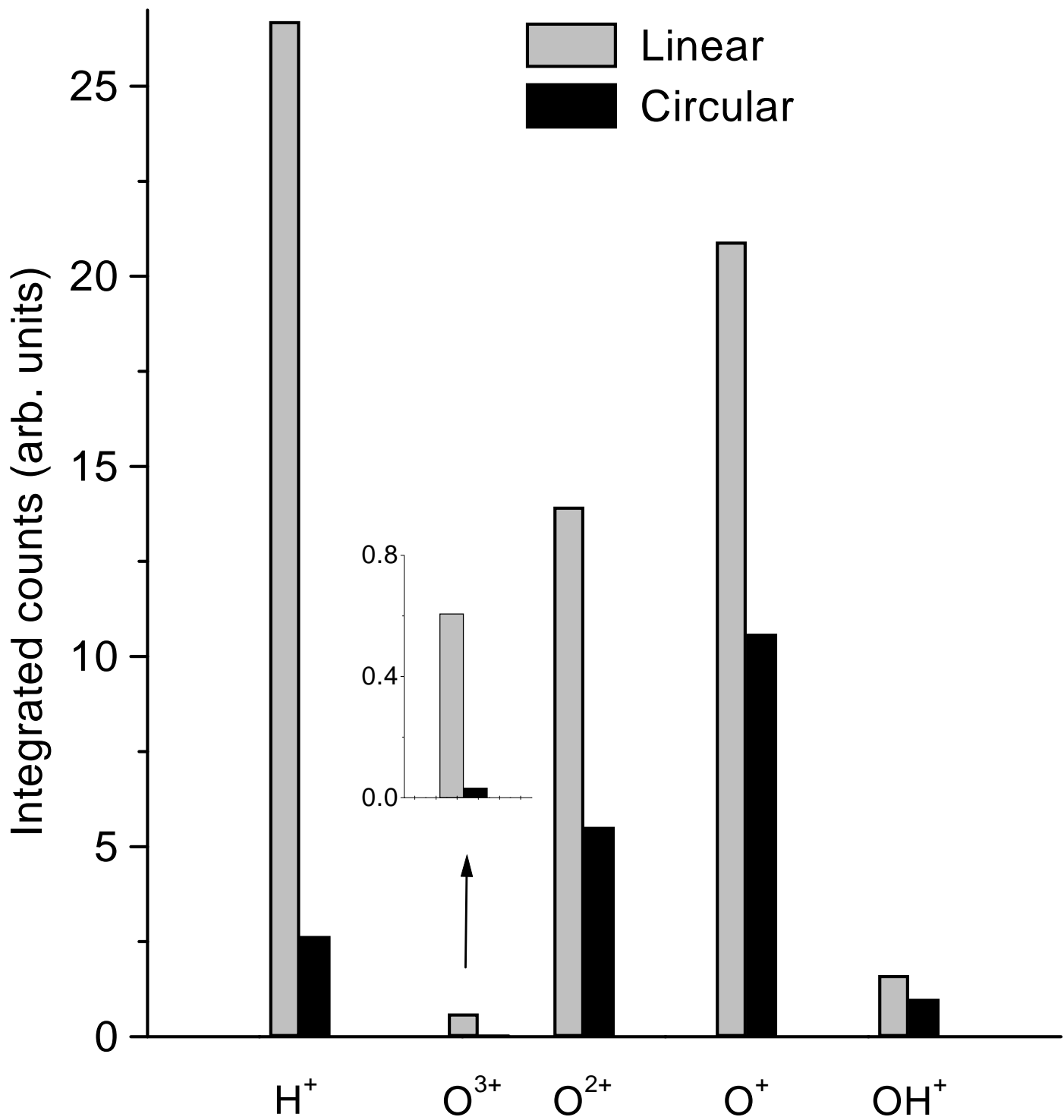
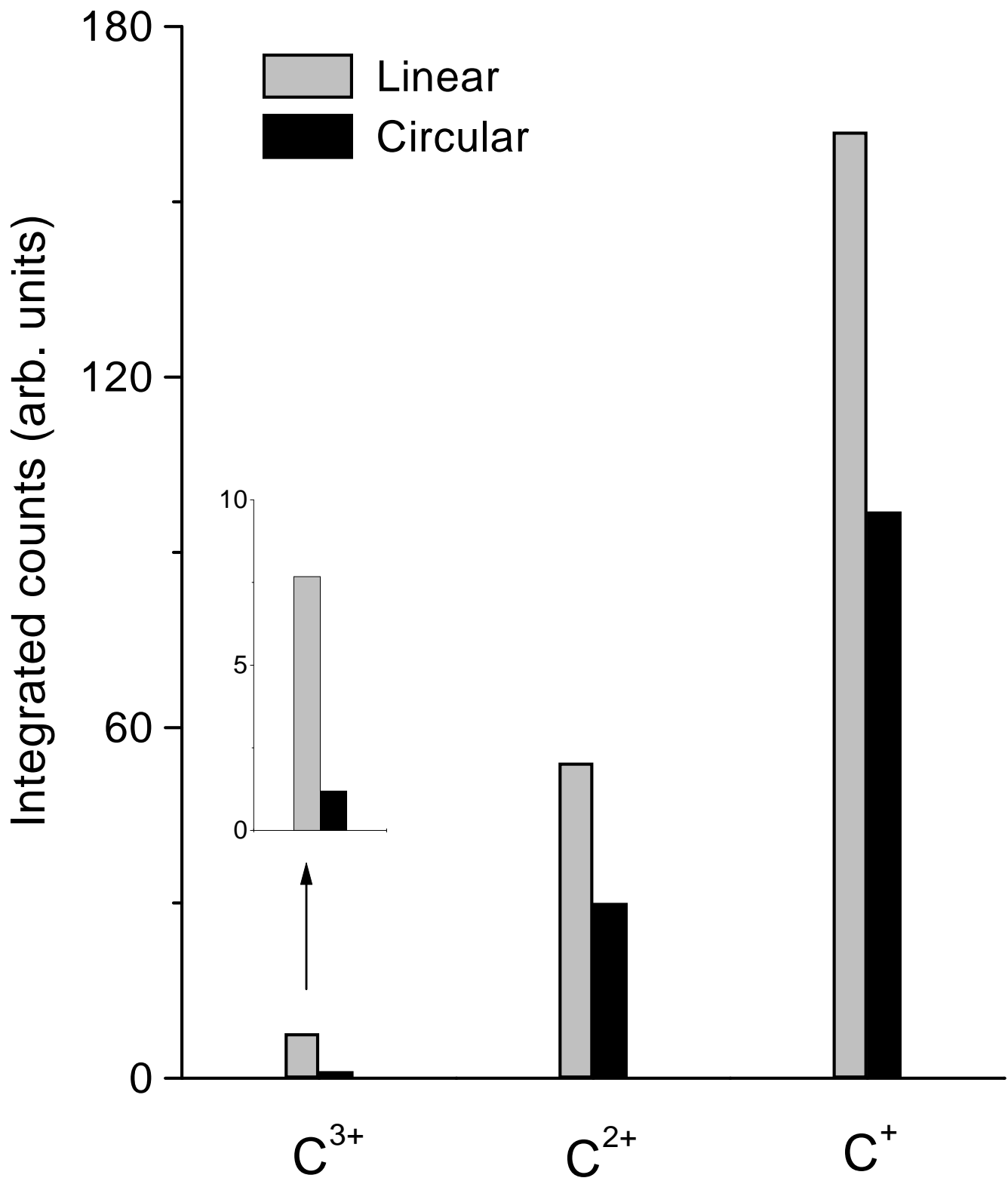
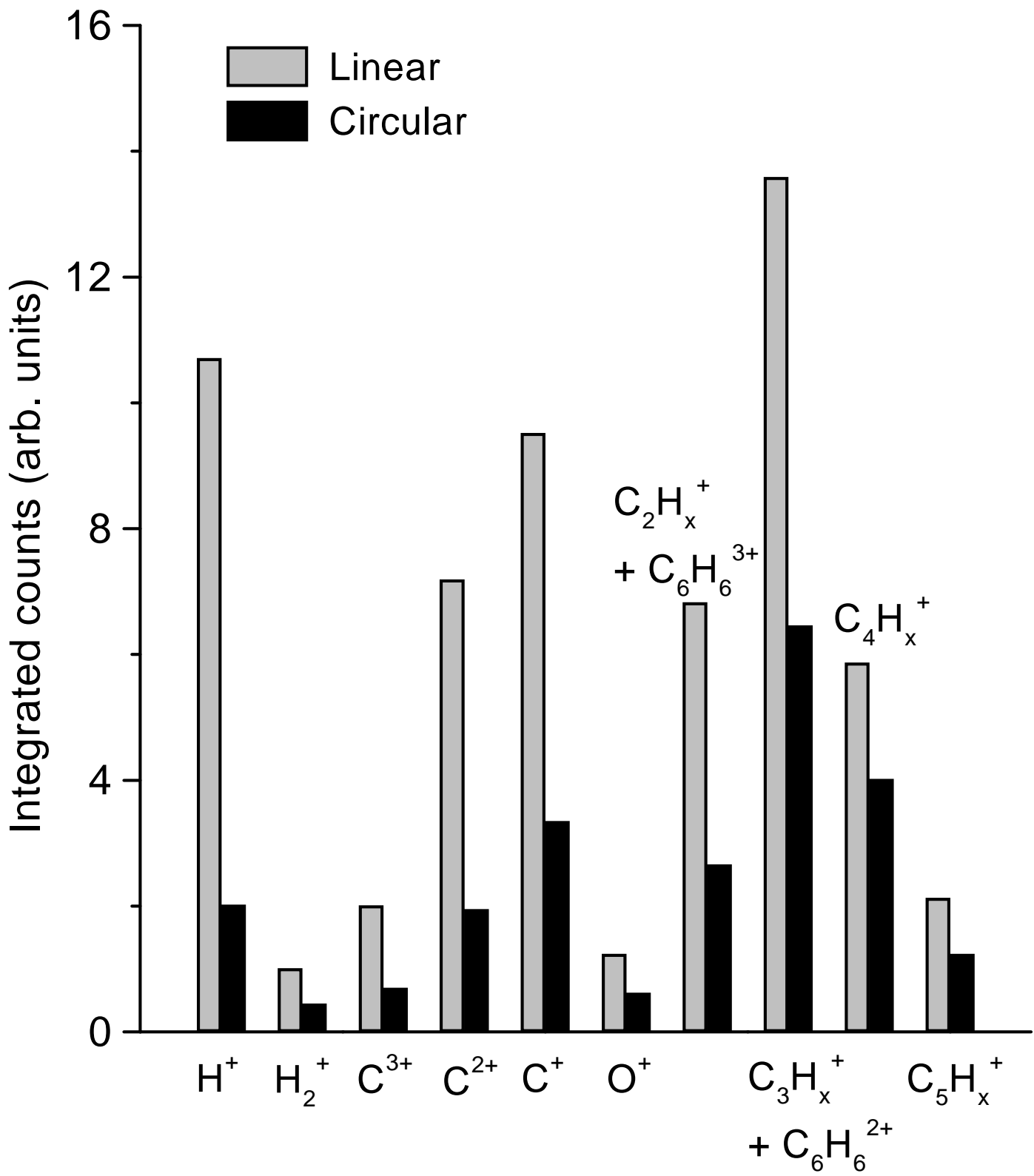
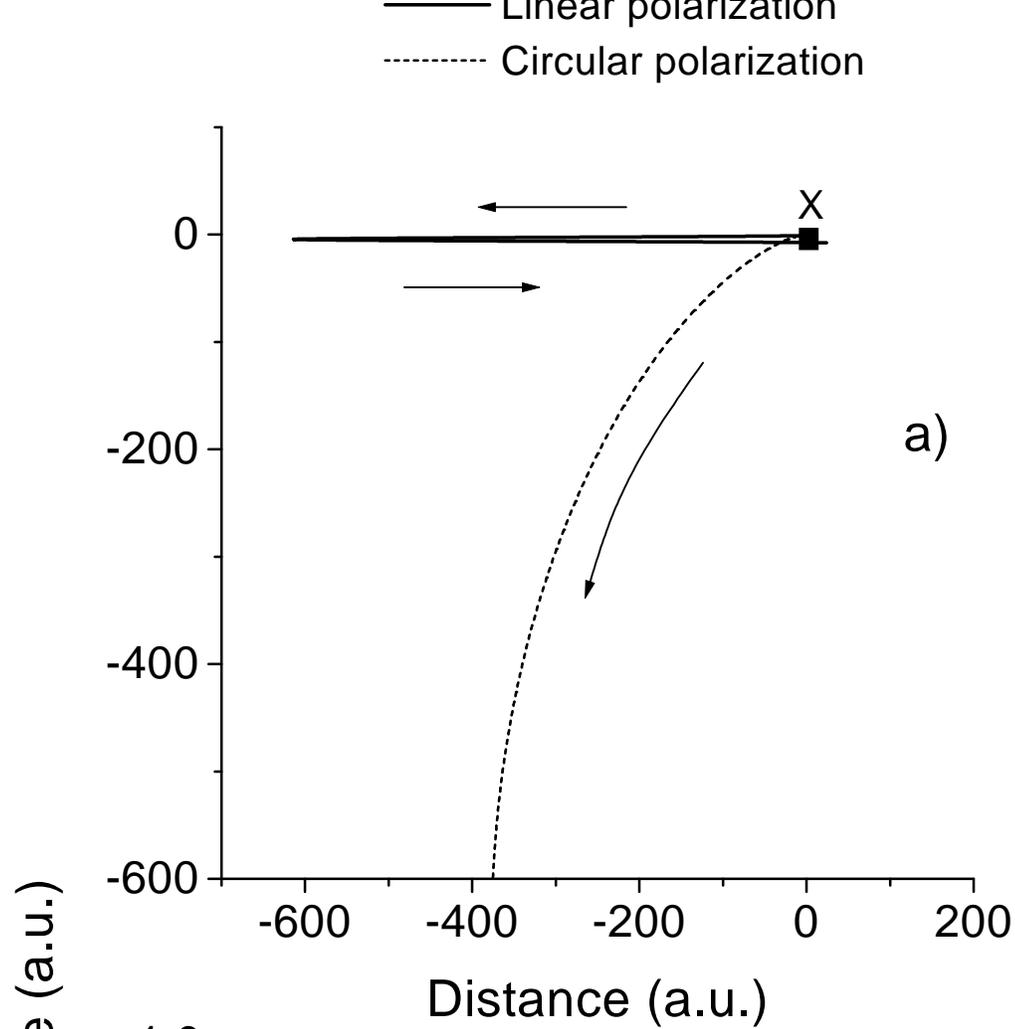
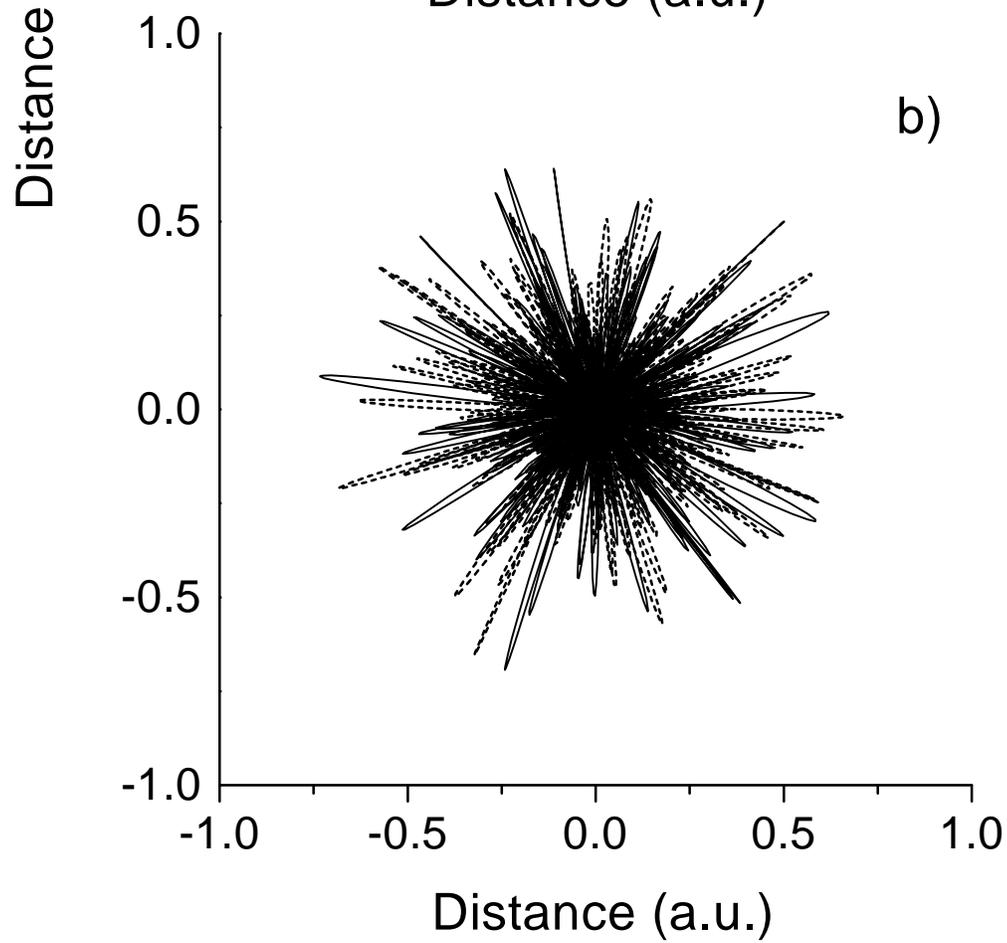
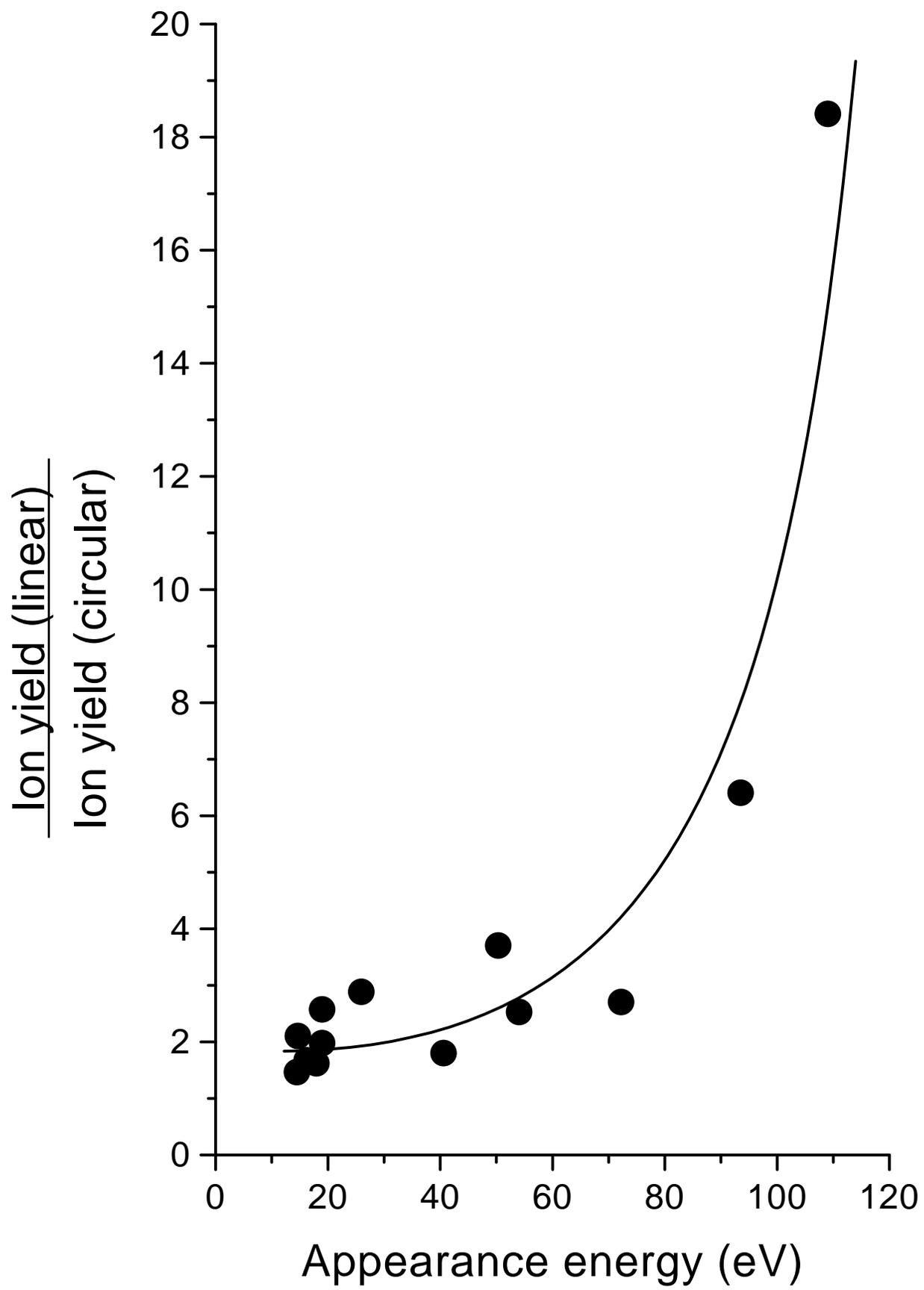
\end{document}